\documentstyle[pra,aps,twocolumn,psfig]{revtex}
\draft
\def\bef{\begin{figure}}
\def\eef{\end{figure}}
\def\bet{\begin{table}}
\def\eet{\end{table}}

\begin{document}
\title{Experimental Investigations of Signals of a New Nature \\
with the aid of two High Precision Stationary Quartz Gravimeters.}

\author{ Yu.A.BAUROV\thanks{baurov@www.com}}
\address{ Central Research Institute of Machine Building\\
141070, Moscow region, Korolev, Pionerskaya, 4, Russia}

\author{ A.V.KOPAJEV}
\address{ Shternberg State Astronomical Institute of 
Moscow State University (SSAI MSU)\\
119899, Moscow, University prospect, 13, Russia}
\maketitle

\begin{abstract}
In consequence of long-term (2000) observation of a system of two
high-precision quartz gravimeters (one of them with an attached magnet)
placed in a special (at a depth of $\sim $10m) gravimetric laboratory on a
common base separated from the foundation of the building, signals of a new
nature were detected. They were of a smooth peak-type shape, several minutes
duration, and with amplitudes often more than that of the moon tide.

The nature of the signals cannot be explained in the framework of
traditional physical views but can be qualitatively described with the aid
of a supposed new interaction connected with the hypothesis about the
existence of the cosmological vectorial potential {\bf A}$_{{\rm g}}$, a
new presumed fundamental vectorial constant.
\end{abstract}

\pacs{PACS numbers: 04.80.+Z, 11.10.Lm, 98.80.-k}

\section{Introduction.}

The signal of a new nature were first detected in a run of experiments of
1994-1996, carried out with the aid of a modernized high-precision quartz
gravimeter ``Sodin'' and a magnet attached to it [1-4]. The signal appeared
as permanent smooth peak-shape pulses with duration from 2 to 10min. and
with amplitudes up to $\sim $ 15L (L is the amplitude of the moon tide). The
experiments were carried out within a special box of the gravimetric
laboratory of SSAI MSU, in a basement room at a level of 2 meters
underground. The duration of the uninterrupted experiment was no more than
30 days. Besides the last peak (19.04.96), all results were obtained on the
basis of one gravimeter measurements. In the last experiment two gravimeters
were used. They were positioned on the same base in the box: one of the them
was with a magnet, the other (reference one) was without magnet. Both
gravimeters were connected to the same power network but had separated
stabilized power sources.

In the experiment with two gravimeters, that with magnet detected an
anomalous event which was not fixed by the gravimeter without magnet.

Since the procedure of investigating the signal of new nature with the aid
of two gravimeters was of short-term duration (and the main results were
obtained by one gravimeter), the run of experiments in 1994-1996 could be
considered only as preliminary one. Nontheless, it should be noted that the
signals (12 peaks) were concentrated in space around some stellar direction
having the following coordinates in the second equatorial system: right
ascension $\alpha\approx270^\circ$ , declination $\delta\approx34^\circ$. This was nearly coincident with
some chosen spatial direction found in the course of investigating the
assumed new anisotropic interaction with the aid of a torsion balance
arranged inside high-current magnets [3-8] as well as when observing time
changes in the intensity of the $\beta$-decay of radioactive elements [3,4,9-10]
in the process of their rotation together with the Earth.

The new run of experiments was aimed at more precise measurements owing to
elimination of technique errors of the preceding investigations.

\section{The experimental setup and procedure.}

In the run of experiments of 1999-2000, the method of two gravimeters with a
magnet attached to one of them, was used. The gravimeters were positioned in
an underground gravimetric laboratory of SSAI MSU, at a depth of $\sim $
10m, on a special base separated from the foundation of the building. A
recently made stabilized power system could feed the gravimeters also in a
case of accidental power switching off during $\sim $ 2hs. The information
from gravimeters was transmitted to a personal computers (PC).

A schematic diagram of a quartz sensitive system is shown in Fig.1. Its main
component is a quartz lever 1 (2 cm in length with a platinum mass of {\it m = 0.05g})
suspended on torsion quartz fibres 2 and additionally off-loaded by a
vertical quartz spring 3. Such a construction gives to the device the high
sensitivity to changes in gravity up to 1{\it $\mu$Gal} ($10^{ - 8}${\it ms}$^{ - 2} = 10^{ - 9}G, G$ is
the free fall acceleration) as well as sufficient protection against
microseismic disturbances.

The optical recording system comprises a galogen lamp 4 fed from a
high-stabilised power source, light of which lamp enters into the instrument
through the objective lens 5 with the aid of optical fibers. The quartz rod 6
welded to the level is a cylindrical lens, it forms an image of the point
light source, which falls further on a photosensor rule 7 with a sensitivity
on the order of 1$\mu$m. The digital signal from the photosensor rule output enters
further through a special interface to a computer PC which executes
preliminary processing and averaging of data over the interval of 1 min. The
accuracy of minute's data is $(0.5-1.2)\times10^{- 9}G$ depending on the microseismic noise
level varying considerably in the course of day.

The sensitive system of the instrument includes some additional units (not
shown in Fig.1) protecting it from thermal, electrostatic, and atmospheric
pressure disturbances.

Calibration of the instrument is carried out by the inclination procedure
with an accuracy of about $(0.5-1.0)\%$ which is quite sufficient for the experiment
being considered.

The amplitude of constantly recorded by the device changes in gravity due to
Moon-Sun tides and corresponding Earth's deformation, is $(50-200)\times10^{-9}G$ which
allows to evaluate magnitudes of possible anomalous effects against the
background of tides.

To measure the new interaction by the Sodin gravimeter, a constant magnet
(60 mm in diameter, $15 $mm in height, the field B in the centre of $0.3T$) was
attached to it in such a way that the vector potential lines of the magnet
in the vicinity of a test platinum weight (see Fig.1) were directed
perpendicular to the Earth's surface. The magnet played the role of a
peculiar amplifier of the new force (see in detail Refs. [3,4]).

\section{Results of experimental investigations.}

To illustrate the experiment, in Figs.2-6 some temporal fragments of
recording of the gravimeters are shown. They demonstrate smooth peak-form
overshoots by the ``Sodin'' gravimeter S209 with the attached constant
magnet.

At the first stage of experiments, the gravimeter with the magnet and the
reference gravimeter worked on different time scales (Moscow (MSK) and
Greenwich (UTC) time, respectively). Therefore a shift of the curves is
seen in Figs.2,3 which corresponds to that of time zones (i.e. 3hs).

It is seen from the Figures that the amplitudes of overshoots are in some
cases (Figs.4,5) much superior to the amplitude of the moon tide. In
February-March 2000, the overshoots corresponded to an increase in the
gravitational effect, in May-June on the contrary to its decrease.

As opposed to the experiments of 1994-1996 in which the duration of the
peaks was various (from 2 to 10 min. and more), the peak length in
consideration was practically constant and equaled 2-3 min.

In Table 1 presented are all anomalous deflections recorded by the
gravimeter with magnet and not fixed by the reference gravimeter without
magnet.

Consider in detail an experimental fragment shown in Fig.6. As is seen, the
events 1,3,4 corresponding to local earthquakes are neatly coincident in the
recordings of both gravimeters, but the event 2 was fixed only by the
gravimeter with magnet.

\section{Discussion.}

As was said above, the instruments in the course of the experiment were
arranged on the base separated from the foundation of the building, at a
depth of $\sim $ 10m. Therefore, as is clearly seen from Fig.6, various
Earth's oscillations acted identically onto both gravimeters.

Their design gives them good protection against thermal, electrostatic, and
barometric disturbances. The gravimeters were powered from the same network,
therefore, if some disturbance of ``unknown nature''\footnote{ The electric
power supply of the instrumentation was stabilized, so that there was no
problem with it and it was not switched off.} could pass through it that
would manifest itself in both readings. The possible ``jumps'' in quartz
occurring sometimes when using the gravimeters ``Sodin'' have quite
different form and cannot explain the nature of the smooth minute's peaks
fixed in the experiment. The result obtained also cannot be explained by
noise in the computers, because during the long use of them before the run
of experiments of 1994-1996 as well as in other experiments carried out in
2001, such peaks were not observed. The outer electromagnetic noise was
practically minimum and, which is the main, that would uniformly tell on
both gravimeters.

One cannot also explain the obtained results by the influence on the
gravimeter of various hypothetical particles predicted in modern calibration
theories (like axions, photinos, gravitinos, etc.) [11,12] since both
gravimeters are equal for their properties.

A temporal analysis of peak's appearance in readings of the gravimeter in
association with the solar activity has shown the following (See Table 1).

In Table 1 shown are the times of detection of the peaks by the gravimeter
with magnet during the first half-year 2000 as well as times of
corresponding events in the Sun with indication of character of the event
and the observatory from which the information was obtained. The more
complete information about the solar events accompanying the peaks can be
found in Refs. [13].

As is seen from Table 1, the peak detected 24.03.2000 by the gravimeter with
magnet coincides practically precisely with a sweep-frequency radio burst.

The event (2) in Fig.6 occurred 5 minutes before a potent solar flare. The
flares of such potency were not observed earlier during more than 40 years.

The other events fixed by the gravimeter lie several minutes or tens of
minutes earlier or later than solar events of ordinary importance.

The results obtained can be qualitatively explained with the help of assumed
new interaction caused by the existence of the vector {\bf A}$_{\rm g}$.

According to recent investigations of anisotropic properties of the new
force presented in Ref.[14], this force is directed at each point of space
over a cone with the opening of $90-100^\circ$ formed around the vector
{\bf A}$_{{\rm g}}$ with the coordinates $\alpha\approx293^{{\circ}}\pm10^{\circ};
\delta\approx36^{{\circ}}\pm10^{\circ}.$ The results of [14] are in good
agreement with [15].

The force can act also in direction of {\bf A}$_{{\rm g}}$ if an object
being acted upon has a closed circular current. This was observed in
experiments on investigating changes in the intensity of $\beta$-decay of
radioactive elements [15].

An analysis of the experimental material presented here has shown that all
events in Table 1 (except (2) in Fig.6) corresponded to such observation
time when the direction of a normal to the surface of the Earth (i.e. the
direction of the maximum sensitivity of the gravimeters) was coincident (to
a precision of $\pm10^\circ$) with the generator of above said cone. The time of the
event (2) in Fig.6 indicates the direction of the normal to the Earth's
surface coincident (to $\pm5^\circ$) with the coordinates of the vector
{\bf A}$_{\rm g}$ itself: $\alpha = 293^\circ$, $\delta = 36^\circ$
(i.e. with the axis of the cone).

It was shown in Ref.[16] that the observed anisotropy of solar flares
reveals a very similar direction ($\alpha = 277^\circ\pm5^\circ$, $\delta = 38^\circ\pm5^\circ$) which can be
explained on the basis of the new anisotropic interaction.

Thus the experiments carried out with the use of two gravimeters have
confirmed the earlier results obtained with the aid of only one gravimeter
with an attached magnet. Their results (the direction of the new force) are
coincident with those of the experiments on investigating the anisotropic
properties of space with the aid of plasma generators as well as the changes
in intensity of the $\beta$-decay of radioactive elements.

The authors are grateful to participants of the seminar at the Astrocosmic
center of the Physical Institute of RAS and personally to prof.
V.V.Burdyuzsa for a fruitful discussion of the experimental results, as well
as to E.P.Morozov, L.I.Kazinova, A.Yu.Baurov for the help in preparation of
the text of the paper.



\draft

\newcommand{\PreserveBackslash}[1]{\let\temp=\\#1\let\\=\temp}
\let\PBS=\PreserveBackslash
\begin{table}
\begin{tabular}
{c|p{42pt}|p{35pt}|p{35pt}|p{35pt}|p{30pt}|p{30pt}}
\raisebox{-1.0ex}[0cm][0cm]{N}&
\raisebox{-1.0ex}[0cm][0cm]{{\bf S209  
}}&
\multicolumn{5}{p{165pt}}{ \bf \hspace{35pt}The events in the Sun}  \\
\cline{3-7}
 &
{\bf events}&
{\bf Begin}&
{\bf Max}&
{\bf End}&
{\bf Obs.}&
{\bf Type} \\
\hline
1.&
07.02.2000 \par 09$^{{\rm h}}$&
08$^{{\rm h}}$54$^{{\rm m}}$ \par 09$^{{\rm h}}$42$^{{\rm m}}$&
/ / / / \par 09$^{{\rm h}}$50$^{{\rm m}}$&
08$^{{\rm h}}$55$^{{\rm m}}$ \par 09$^{{\rm h}}$57$^{{\rm m}}$&
LEA \par LEA&
RSP \par FLA \\
\hline
2.&
04.03.2000 \par 06$^{{\rm h}}$05$^{{\rm m}}$&
05$^{{\rm h}}$41$^{{\rm m}}$ \par 06$^{{\rm h}}$20$^{{\rm m}}$&
05$^{{\rm h}}$42$^{{\rm m}}$ \par 06$^{{\rm h}}$28$^{{\rm m}}$&
05$^{{\rm h}}$57$^{{\rm m}}$ \par 07$^{{\rm h}}$08$^{{\rm m}}$&
LEA \par LEA&
FLA \par FLA \\
\hline
3.&
24.03.2000 \par 01$^{{\rm h}}$08$^{{\rm m}}$&
00$^{{\rm h}}$35$^{{\rm m}}$ \par 01$^{{\rm h}}$08$^{{\rm m}}$ \par 03$^{{\rm h}}$09$^{{\rm m}}$&
00$^{{\rm h}}$40$^{{\rm m}}$ \par / / / / \par 03$^{{\rm h}}$14$^{{\rm m}}$&
00$^{{\rm h}}$47m \par 01$^{{\rm h}}$09$^{{\rm m}}$ \par 03$^{{\rm h}}$17$^{{\rm m}}$&
GO8 \par LEA \par GO8&
XRA \par RSP \par XRA \\
\hline
4.&
26.05.2000 \par 17$^{{\rm h}}$05$^{{\rm m}}$&
16$^{{\rm h}}$25$^{{\rm m}}$ \par 18$^{{\rm h}}$02$^{{\rm m}}$&
16$^{{\rm h}}$26$^{{\rm m}}$ \par 18$^{{\rm h}}$07$^{{\rm m}}$&
16$^{{\rm h}}$29$^{{\rm m}}$ \par 18$^{{\rm h}}$16&
HOL \par GO8&
FLA \par XRA \\
\hline
5.&
27.05.2000 \par 15$^{{\rm h}}$36$^{{\rm m}}$&
14$^{{\rm h}}$45$^{{\rm m}}$ \par 15$^{{\rm h}}$49$^{{\rm m}}$&
/ / / / \par 15$^{{\rm h}}$50$^{{\rm m}}$&
14$^{{\rm h}}$45$^{{\rm m}}$ \par 16$^{{\rm h}}$00$^{{\rm m}}$&
SVI \par RAM&
RSP \par FLA \\
\hline
6.&
06.06.2000 \par 12$^{{\rm h}}$01$^{{\rm m}}$&
12$^{{\rm h}}$00$^{{\rm m}}$ \par 12$^{{\rm h}}$06$^{{\rm m}}$ \par 12$^{{\rm h}}$08$^{{\rm m}}$&
12$^{{\rm h}}$18$^{{\rm m}}$ \par 15$^{{\rm h}}$21$^{{\rm m}}$ \par 12$^{{\rm h}}$12$^{{\rm m}}$&
12$^{{\rm h}}$25$^{{\rm m}}$ \par 18$^{{\rm h}}$43$^{{\rm m}}$ \par 12$^{{\rm h}}$13$^{{\rm m}}$&
GO8 \par RAM \par SVI&
XRA \par FLA \par RBR \\
\end{tabular}
\vskip10pt
{\bf Table 1.} The events were recorded by the gravimeter with magnet and
the events in the Sun.

{\bf Obs.} (the reporting observatory):
GO8 - GOES-8 satellite;
HOL - Holloman AFB, NM, USA;
SVI - San Vito, Italy;
LEA - Learmonth, Australia;
RAM - Ramey AFB, PR, USA.

{\bf Type} (types of events):
FLA - Optical flare observed in H-alpha;
RBR - Fixed-frequency radio burst;
RSP - Sweep-frequency radio burst;
XRA - X-ray flare.

{\bf Begin, Max, End} - the UTC time of the beginning, maximum, and
the end of the event, //// - indicates a missing time.
\end{table}
\draft

\draft
\bef
\protect\hbox{\psfig{file=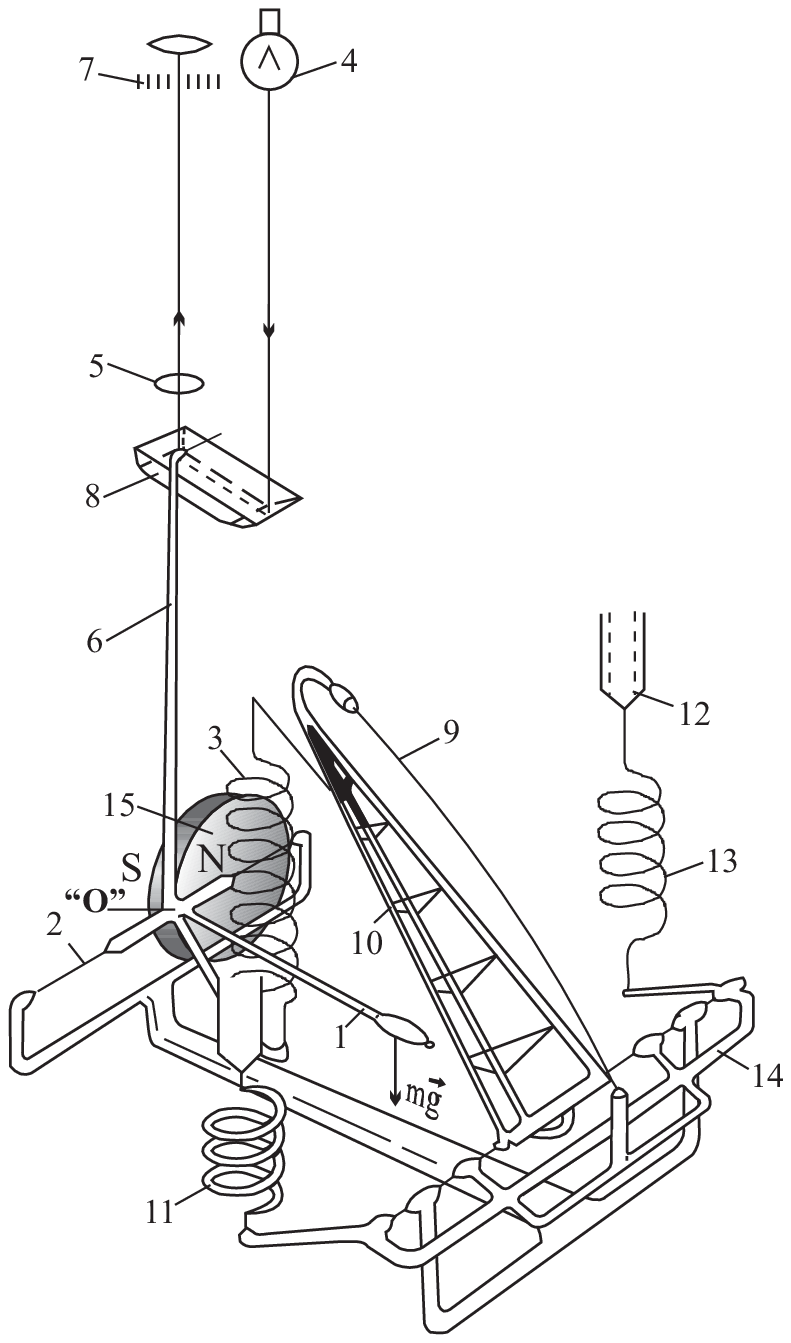,height=16.6cm,width=9.0cm}}
\label{1}
Fig.1 Sensitive system of the Sodin quartz gravimeter:
 1   -	 beam, \\
2   -   horizontal quartz wires	 suspension system, \\
3   -   main spring, \\
4   -	 lamp, \\
5   -   object lens, \\
6   -   beam, \\
7   -	 CCD-SCALE,\\
 8   -   prism, \\
9, 10  -   thermocompensator, \\
11, 12, 13, 14 - micrometric compensation mechanism, \\
15  -  constant magnet.
\eef
\draft

\draft
\vspace*{2cm}
\bef
\protect\hbox{\psfig{file=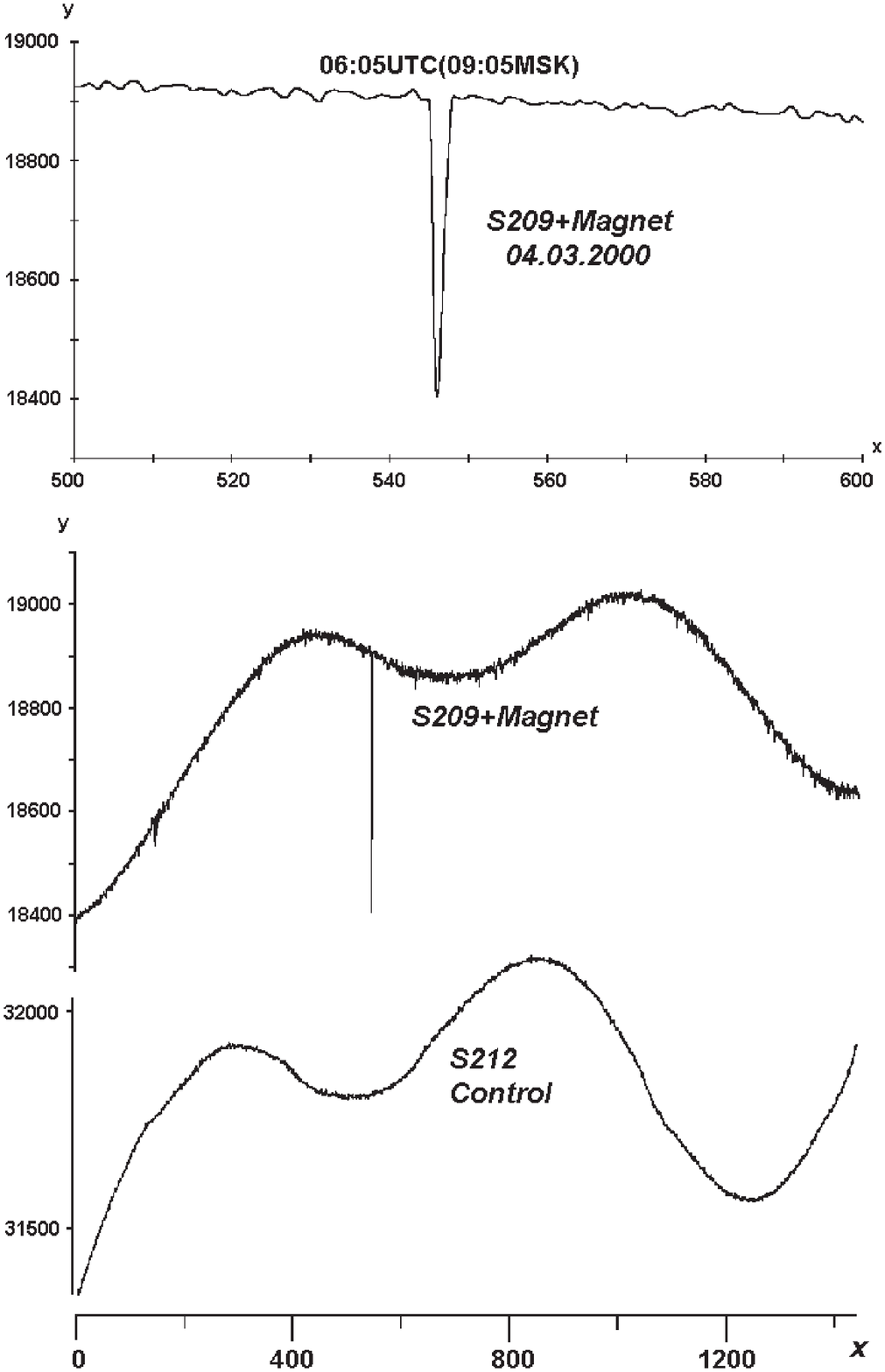,height=13.5cm,width=9.0cm}}
\label{2}
Fig.2 Readings of the gravimeters at Mar.04.2000.\\
  $y$  -  the displacement of platinum weight
(one division is equal to $0.1\mu$ and corresponds to $0.2\mu Gal$); \\
$x$  -  time (in minutes);

\parindent35pt
Event (peak) at $6^h5^m$ UTC ($9^h5^m$ MSK).
\eef

\draft
\vspace*{2cm}
\bef
\protect\hbox{\psfig{file=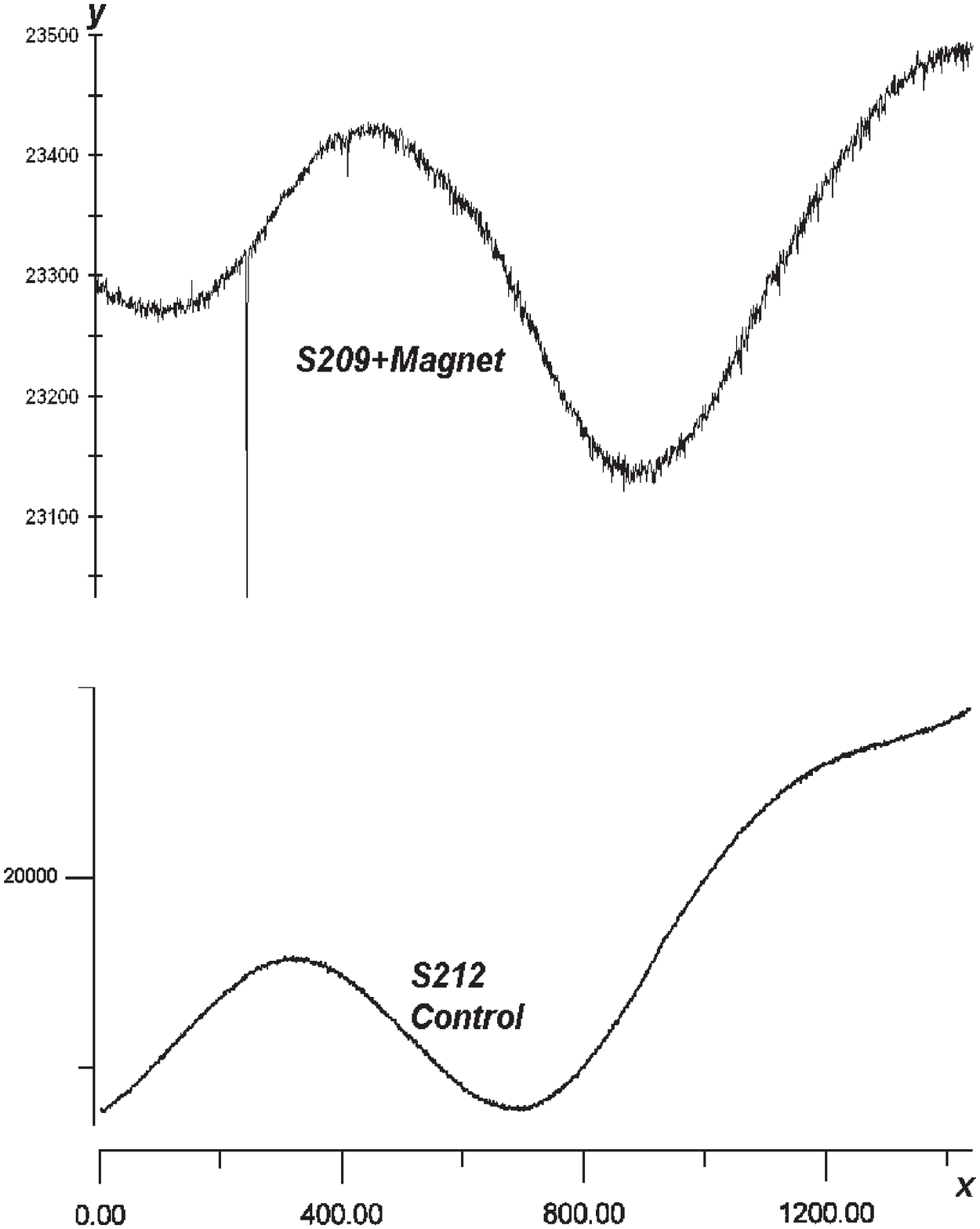,height=12.7cm,width=9.0cm}}
\label{3}
\vskip10pt
Fig.3 The same as in Fig.2 at Mar.24.2000.

\parindent35pt
Event (peak) at $1^h8^m$ UTC ($4^h8^m$ MSK).
\eef
\pagebreak
\vspace*{2cm}
\draft
\bef
\protect\hbox{\psfig{file=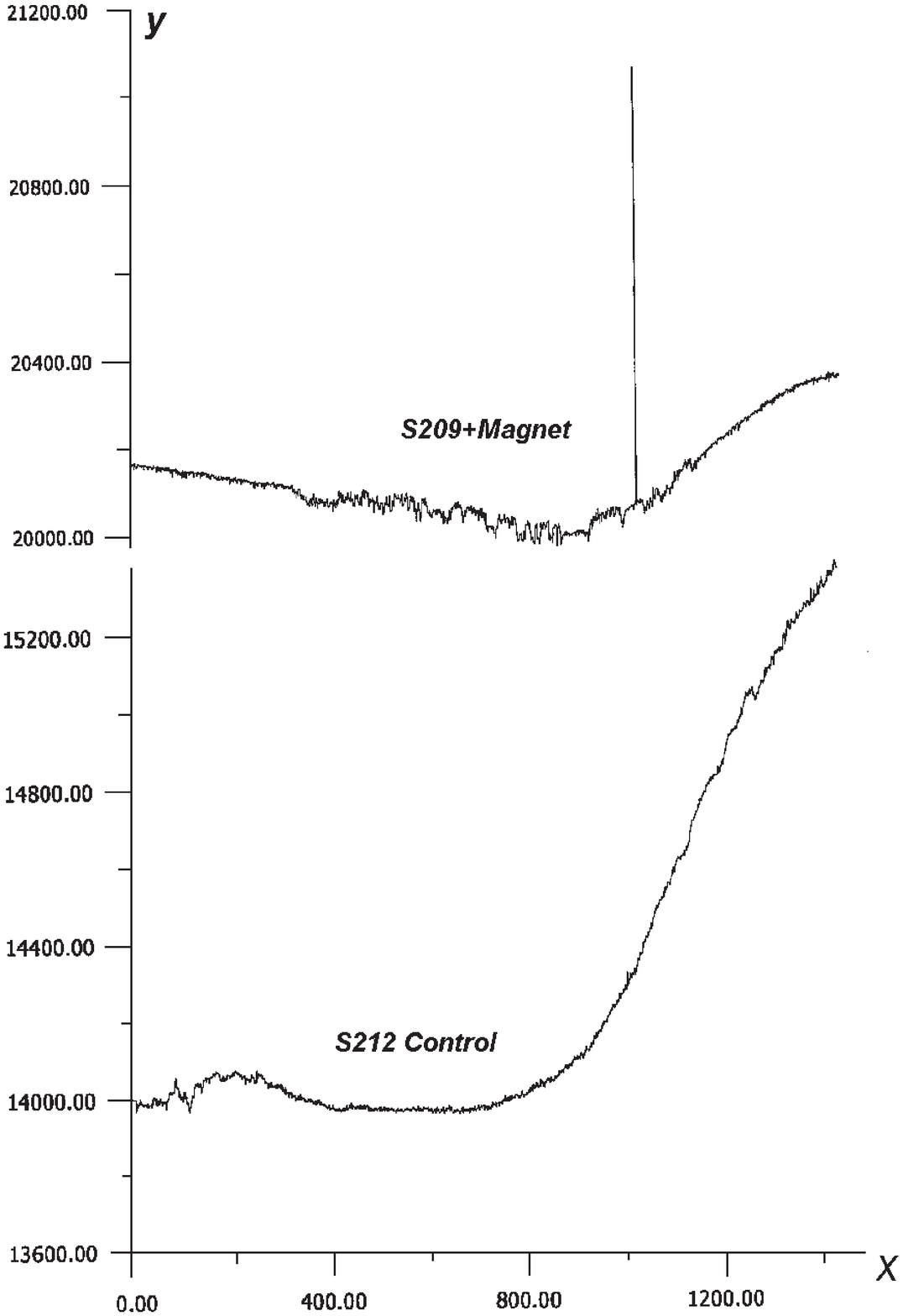,height=12.7cm,width=9.0cm}}
\label{4}

Fig.4 The same as in Fig.2 at May.26.2000.

\parindent35pt
Event (peak) at $17^h5^m$ UTC.
\vskip10pt
\eef
\draft
\vspace*{2cm}
\bef
\protect\hbox{\psfig{file=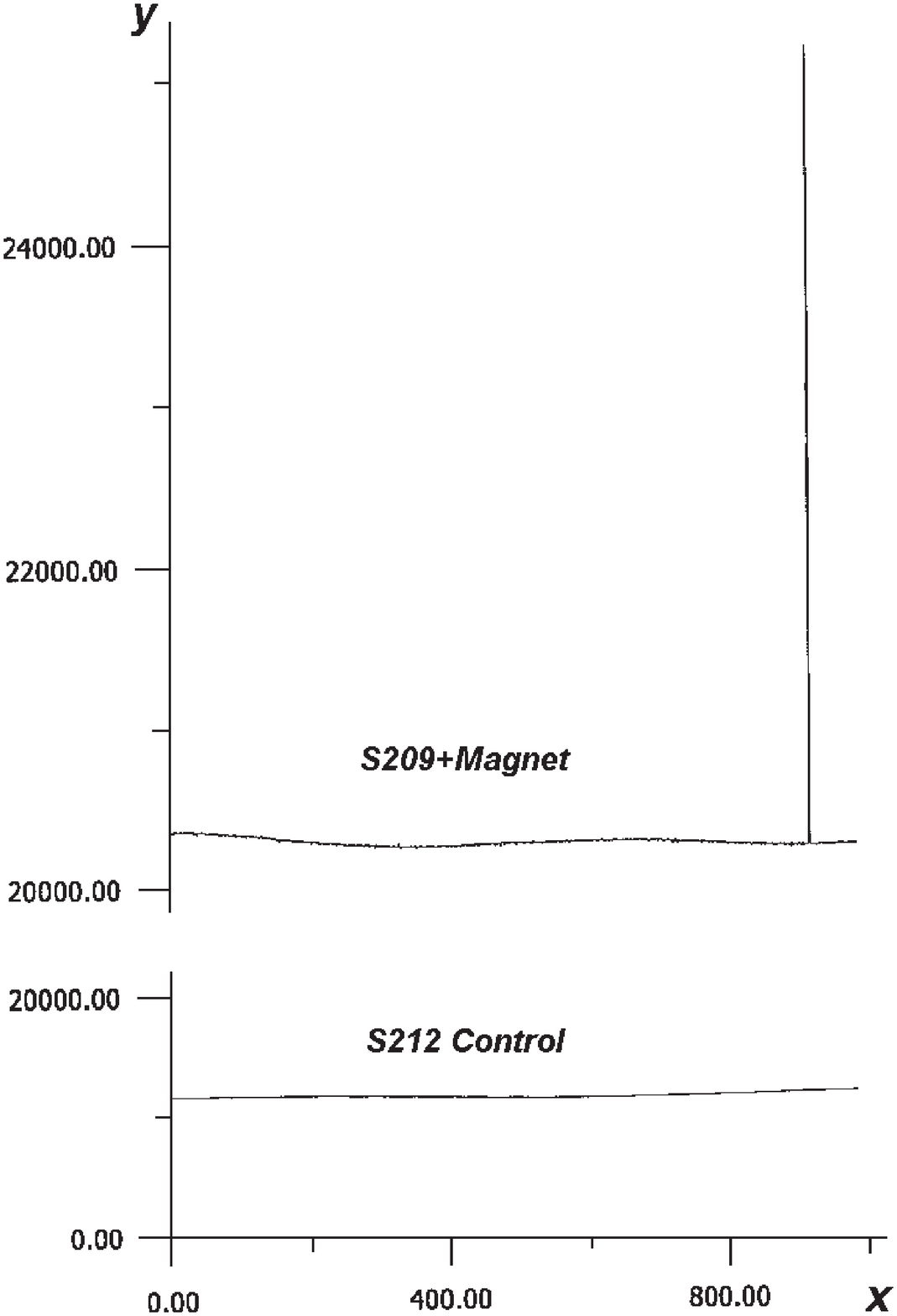,height=12.7cm,width=9.0cm}}
\label{5}

Fig.5 The same as in Fig.2 at May.27.2000.

\parindent35pt
Event (peak) at $15^h36^m$ UTC.
\vskip10pt
\eef
\draft
\vspace*{2cm}
\bef
\protect\hbox{\psfig{file=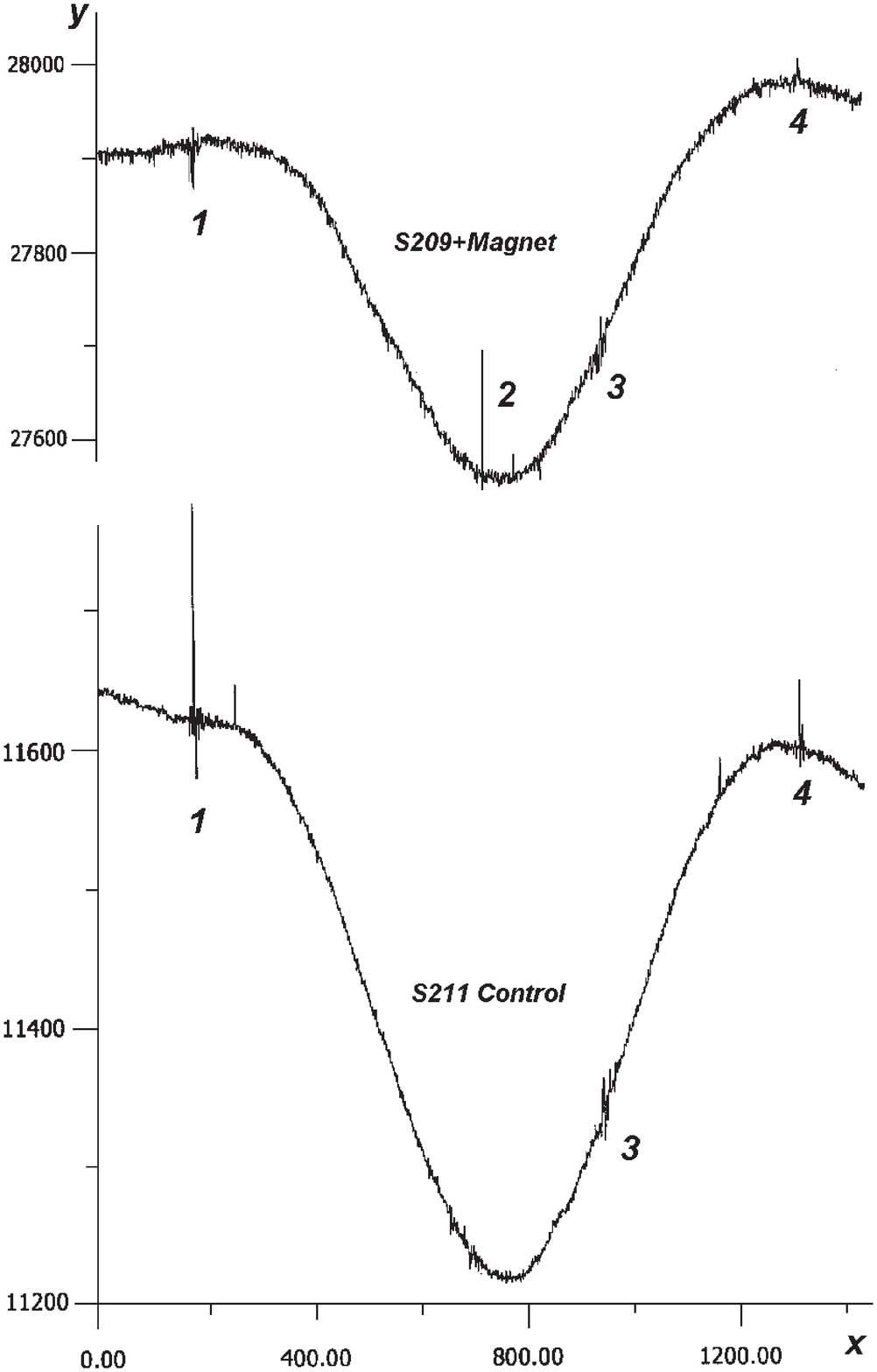,height=12.7cm,width=9.0cm}}
\label{6}
Fig.6 The same as in Fig.2 at Jun.6.2000.

\parindent35pt
Event (peak) at $12^h1^m$ UTC.
\vskip10pt
\eef
\end{document}